\documentclass[11pt]{article}

\usepackage{fullpage}
\usepackage{authblk}
\usepackage{natbib}
\usepackage{algorithm}
\usepackage{algpseudocode}
\usepackage{amsmath,amsthm,amsfonts}
\usepackage{amsmath}
\usepackage[subnum]{cases}
\usepackage{tcolorbox}
\usepackage{mathrsfs}
\usepackage{amssymb}
\usepackage{color}
\usepackage{mathrsfs}
\usepackage{enumitem}
\usepackage{bm}
\usepackage{multirow}
\usepackage{booktabs}
\usepackage{makecell}
\usepackage{graphicx}
\usepackage{subcaption}
\usepackage{comment}
\usepackage{cases}
\usepackage{appendix}
\usepackage{tikz}
\usetikzlibrary{arrows,shapes}
\usepackage[colorlinks= true, linkcolor=red, citecolor=blue, urlcolor=black]{hyperref}
\usepackage{cleveref}
\usepackage{marginnote}
\usepackage{diagbox} 

\numberwithin{equation}{section}

\theoremstyle{definition}

\usepackage{booktabs}
\usepackage{tikz}
\usepackage{pgfplots}
\usetikzlibrary{calc,intersections}

\usetikzlibrary{shapes.geometric, arrows, positioning}

\tikzset{
	mybox/.style  = {draw, rectangle, minimum width=4cm, minimum height=0.8cm, text centered, text width=4.4cm,   
		font=\normalsize},
	box/.style  = {draw, rectangle, minimum width=2.0cm, minimum height=0.6cm, text centered, text width=3.0cm,   
		font=\normalsize},
	myarrow/.style = {line width=0.2pt, draw=black, -triangle 60, postaction={draw, line width=0.2pt, shorten >=10pt,-}}
}

\tikzstyle{arrow} = [->, >=stealth, -triangle 60]

\allowdisplaybreaks

\makeatletter
\newcommand{\leqnomode}{\tagsleft@true}
\newcommand{\reqnomode}{\tagsleft@false}
\makeatother

\begin{document}

\title{The Sampling Method for Optimal Precursors of ENSO Events}

\author[1,2]{Bin Shi\thanks{Corresponding author: \url{shibin@lsec.cc.ac.cn} } }
\author[3]{Junjie Ma\thanks{Junjie Ma joins in this work during the final semester of his Ph.D. under the supervision of Wansuo Duan at the Institute of atmospheric physics, Chinese Academy of Sciences.}}
\affil[1]{State Key Laboratory of Scientific and Engineering Computing, Academy of Mathematics and Systems Science, Chinese Academy of Sciences, Beijing 100190, China}
\affil[2]{School of Mathematical Sciences, University of Chinese Academy of Sciences, Beijing 100049, China}
\affil[3]{School of Mathematics, North University of China, Taiyuan 030051, China}


\date\today

\maketitle

\begin{abstract}
El Ni\~{n}o-Southern Oscillation (ENSO) is one of the significant climate phenomena, which appears periodically in the tropic Pacific.  The intermediate coupled ocean-atmosphere Zebiak-Cane (ZC) model is the first and classical one designed to numerically forecast the ENSO events. Traditionally, the conditional nonlinear optimal perturbation (CNOP) approach has been used to capture optimal precursors in practice. In this paper,  based on state-of-the-art statistical machine learning techniques,\footnote{Generally, the statistical machine learning techniques refer to the marriage of traditional optimization methods and statistical methods, or, say, stochastic optimization methods, where the iterative behavior is governed by the distribution instead of the point due to the attention of noise. Here, the sampling algorithm used in this paper is to numerically implement the stochastic gradient descent method, which takes the sample average to obtain the inaccurate gradient.} we investigate the sampling algorithm proposed in~\citep{npg-30-263-2023} to obtain optimal precursors via the CNOP approach in the ZC model. For the ZC model, or more generally, the numerical models with dimension $\mathrm{O}(10^4-10^5)$, the numerical performance, regardless of the statically spatial patterns and the dynamical nonlinear time evolution behaviors as well as the corresponding quantities and indices, shows the high efficiency of the sampling method by comparison with the traditional adjoint method. The sampling algorithm does not only reduce the gradient (first-order information) to the objective function value (zeroth-order information) but also avoids the use of the adjoint model, which is hard to develop in the coupled ocean-atmosphere models and the parameterization models. In addition, based on the key characteristic that the samples are independently and identically distributed, we can implement the sampling algorithm by parallel computation to shorten the computation time. Meanwhile, we also show in the numerical experiments that the important features of optimal precursors can be still captured even when the number of samples is reduced sharply.

\end{abstract}
%

\section{Introduction}
\label{sec: introduction}

In the global climate system, the most prominent phenomenon of year-to-year fluctuations is El Ni\~{n}o–Southern Oscillation (ENSO), which makes a huge impact on Earth’s ecosystems and human societies via influencing temperature and precipitation~\citep{cashin2017fair}.  The natural interactions between ocean and atmosphere over the tropical Pacific not only alter weather around the world thus affecting marine and terrestrial ecosystems, such as fisheries, but also bring about secondary influences, such as human health and other societal and economic aspects of the Earth system~\citep{mcphaden2006enso, timmermann2018nino, boucharel2021influence}.  Thus,  it is of vital importance to learn the mechanism behind the set of coupled ocean-atmosphere phenomena in order to make a better forecast~\citep{philander1989nino, sarachik2010nino}.

Perhaps the modern studies of the ENSO theory date back to the late sixties of the last century.~\citet{bjerknes1969atmospheric} pioneered the positive feedback mechanism, which explains why the  ENSO system has two favored phases and their rapid growth. However, the positive feedback mechanism does not provide any explanation for why the ENSO event transits between the two phases. For the tropical atmosphere,~\citet{gill1980some} proposed a linear shallow water model on the equator with the first barocline mode vertical structure, which has become a standard tool used by both modelers and diagnosticians for describing the atmospheric response to the equatorially thermal forcing. Afterward, the heating parameterized in terms of sea surface temperature (SST) anomalies with arbitrary distribution was introduced in~\citep{zebiak1982simple} and then convergence feedback parameterization in~\citep{zebiak1986atmospheric}.  For the ocean process in the tropics, the governing equations including reduced gravity upper-ocean momentum equations and continuity equation for ocean thermocline depth were proposed in~\citep{cane1984modeling}. Ultimately,~\citet{zebiak1987model} proposed the milestone coupled ocean-atmosphere model, Zebiak-Cane (ZC) model, to simulate the ENSO event,  which imports the thermodynamical equation in terms of SST anomalies and couples the oceanic motion forced by the wind stress. Although the ZC model simulates the oscillation phenomenon,  the mechanism still remains unclear in~\citep{zebiak1987model}. The so-called delayed oscillator theory proposed in~\citep{suarez1988delayed, battisti1989interannual} introduces the delayed negative feedback for the phase transition, where the core idea is the delayed effect of equatorial ocean waves. Based on the simulation of the ZC model, the well-known recharge-oscillator theory is first developed heuristically by~\citet{jin1997equatorial1, jin1997equatorial2}, where the key process is the zonal mean thermocline variation. Meanwhile, the ZC model is also the first intermediate coupled ocean-atmosphere numerical model used widely for the ENSO forecast. After it was proposed in~\citep{zebiak1987model}, there were many improvements in its predictability. The initialization procedure that incorporates the air-sea coupling was designed by~\citet{chen1995improved}, which substantially improves the predictability of the ZC model. To predict the ENSO event, the ZC model was further improved by assimilating observed sea level data in~\citep{chen1998impact}. The LDEO5 version of the ZC model was exploited in~\citep{chen2004predictability}, which successfully predicts all prominent El Ni\~{n}o events within the period 1857 to 2003 at lead times of up to two years.

In numerical prediction, a key issue that we often meet is the short-time behavior of a predictive model with imperfect initial data. In other words, it is of vital importance to understand the sensitivity of the numerical models to errors in the initial data. The simplest and most practical way is to estimate the likely uncertainty for the initial data polluted by the most dangerous errors. Currently, the conventional approach to capture the optimal initial perturbation is the so-called conditional nonlinear optimal perturbation (CNOP) approach innovatively introduced in~\citep{mu2003conditional}, which is based on nonlinear optimization methods. The CNOPs of the ZC model were first investigated in~\citep{mu2007kind}, which shows that the optimal initial errors probably cause a significant spring predictability barrier (SPB). It is further recognized in~\citep{yu2009dynamics} that there are two kinds of CNOP-type initial errors, a large-scale zonal dipolar pattern for the SST anomalies and a basin-side deepening or shoaling along the equator for the thermocline depth anomalies. Based on the ZC model, it is also revealed in~\citep{mu2014similarities} that there is a great similarity between the optimal precursors and the optimal initial errors obtained by the CNOP approaches in terms of spatial structure and localization. In addition, the ideas based on the CNOP approaches, or more general nonlinear optimization methods,  have been generalized to rectify the model errors on the forecast of ENSO diversity in the ZC model, such as the SST cold-tongue cooling bias condition for the frequent occurrence of the central Pacific (CP)-type El Ni\~{n}o events in~\citep{duan2014simulations} and the nonlinear forcing singular vector (NFSV) perturbation that can distinguish the two kinds of El Ni\~{n}o events, the CP-type El Ni\~{n}o and the east Pacific (EP)-type El Ni\~{n}o in~\citep{tao2020improving}. Furthermore, an ensemble NFSV data assimilation approach is developed to address the ENSO forecast uncertainties caused by the spring predictability barrier and El Ni\~{n}o diversity~\citep{zheng2023using}.   

The CNOPs are often obtained by implementing nonlinear optimization methods, mainly including spectral projected gradient (SPG) method~\citep{birgin2000nonmonotone}, sequential quadratic programming (SQP)~\citep{barclay1998sqp}, and the limited memory Broyden-Fletcher-Goldfarb-Shanno (BFGS) algorithm~\citep{liu1989limited} in practice. As we know, the final state is the nonlinear evolution of the initial data polluted by some dangerous errors via a couple of nonlinear partial differential equations and some more complex parameterization models.  Thus, the direct numerical computation of the gradient is so extremely expensive with the increase of dimension that it is unavailable in practice, since it needs to compute the Jacobian of the final reference state on the initial errors. The most popular and practical way to numerically approximate the gradient is the so-called adjoint technique, where the core is to exploit the adjoint model~\citep{kalnay2003atmospheric}. Generally, the adjoint method reduces the computation time significantly at the cost of massive storage space to save the basic state.  Even though a large amount of storage space has not been an essential issue based on the capabilities of modern computers, the adjoint model is still unusable for many numerical models, since the adjoint models are hard to develop, especially for the coupled ocean-atmosphere models as well as the parameterization models~\citep{wang2020useful}. Based on state-of-the-art statistical machine learning techniques,~\citet{npg-30-263-2023} proposes the sampling algorithm to compute the CNOPs, which is prone to implementation in practice.~\citet{npg-30-263-2023} has successfully shown the efficiency of the sampling algorithm in the theoretical models, such as the Burgers equation with small viscosity and the Lorenz-96 model. Moreover, the computation time is shortened to the utmost at the cost of losing little accuracy. In this paper, we further implement the sampling algorithm to obtain the CNOPs in the realistic and predictive ZC model. Meanwhile, we show the efficiency of the sampling method by comparison with the adjoint method and discuss its available implementation in practice with modern parallel computation techniques. In addition, we also provide a positive answer for the open question of whether there exists an adjoint-free algorithm to obtain the CNOPs directly for the numerical models with dimension $\mathrm{O}(10^4-10^5)$, which has already been listed in~\citep{mu2017nonlinear, wang2020useful}.

The paper is organized as follows.~\Cref{sec: cnop-sampling} briefly describes how to numerically compute optimal precursors of the ENSO events in the ZC model, which includes the basic CNOP settings and the implementation of the sampling algorithm as well as how to carry it out by parallel computation in practice. The numerical performance of the sampling algorithm with the comparison of the adjoint method for the ZC model, in terms of the statically spatial patterns and the dynamical nonlinear time evolution behaviors as well as the corresponding quantities and indices, is shown in~\Cref{sec: numerical-performance}. Finally, we conclude this paper with a brief summary and discussion on some furthre researches in~\Cref{sec: summary-discussion}.

\section{Optimal Precursors via CNOP and Sampling}
\label{sec: cnop-sampling}

In this section, we first briefly describe the basic process to compute the optimal precursors by the use of the CNOP approach in the ZC model.\footnote{Although the CNOP approach has been extended to investigate the influences of boundary errors and model errors on atmospheric and oceanic models~\citep{wang2020useful}, here we only explore the impact of initial errors.} Then, based on the key characteristic that the samples are independently and identically distributed, we point out that the sampling method can be implemented efficiently by parallel computation and provide a detailed discussion.

\subsection{The Basic CNOP Settings}
\label{subsec: cnop}

Let $\pmb{\mathscr{T}} = (\mathscr{T}_{ij})$ and $\pmb{\mathscr{H}}=(\mathscr{H}_{ij})$ be SST anomalies and thermocline depth anomalies respectively,\footnote{Throughout the paper, all the vectors are denoted by the bold italics. } where the index $i$ indicates the longitudinal grids in the region from $129.375\;^{\circ}\text{E}$ to $84.375\;^{\circ}\text{W}$ with the grid space $5.625^{\circ}$ and the index $j$ indicates the latitudinal grids from $19\;^{\circ}\text{S}$ to $19\;^{\circ}\text{N}$ with the grid space $2^{\circ}$. From the classical references~\citep{wang1996chaotic, mu2007kind}, we know that the characteristic scales of the SST anomalies and the thermocline depth anomalies are $|\mathscr{T}| \sim 2\;^{\circ}\text{C}$ and $|\mathscr{H}| \sim 50\;\text{m}$, respectively.  Then, the nondimensionalized quantities of the SST anomalies and the thermocline depth anomalies are given as
\begin{equation}
\label{eqn: sst-td-scale}
\pmb{T} = \frac{\pmb{\mathscr{T}}}{|\mathscr{T}|} = \frac{\pmb{\mathscr{T}}}{2} \quad \text{and} \quad \pmb{H} = \frac{\pmb{\mathscr{H}}}{|\mathscr{H}|}= \frac{\pmb{\mathscr{H}}}{50}.
\end{equation}
Moreover, in the ZC model, the dominant factors that influence the ENSO events are the SST anomalies and the thermocline depth anomalies~\citep{zebiak1987model}. With~\eqref{eqn: sst-td-scale}, the initial errors that we need to consider should include these two variables as $\pmb{u}_0 = (\pmb{T}(0), \pmb{H}(0))$. For the quantity used to measure, we adopt the standard Euclidean norm as
\begin{equation}
\label{eqn: u0-norm}
\|\pmb{u}_0\| = \|(\pmb{T}(0), \pmb{H}(0))\| = \sqrt{\sum_{i,j}\left[T(0)_{ij}^2 + H(0)_{ij}^2\right]}.
\end{equation}

Next, we consider the objective function that is on the initial errors. Since the target quantity required to maximize that we concern is only on the nonlinear evolution state of the SST anomalies, then the objective function is given as
\begin{equation}
\label{eqn: obj-norm}
J(\pmb{u}_0) = \|\pmb{T}(\tau)\|^{2},
\end{equation}
where $\|\cdot\|$ is still the Euclidean norm and $\tau$ is the prediction time set as $9$ months in this paper. With~\eqref{eqn: u0-norm} and~\eqref{eqn: obj-norm}, we derive the constrained nonlinear optimization problems for the optimal precursors, that is, the CNOPs in the ZC model  as
\begin{equation}
\label{eqn: cnop-solve}
\max_{\|\pmb{u}_0\| \leq \delta} J(\pmb{u}_0)
\end{equation}
where the constraint parameter is set as $\delta = 1.0$.

\subsection{The Sampling Method and Parallel Computation}
\label{subsec: parallel-computation}

Based on Stokes' formula,~\citet{npg-30-263-2023} proposes the sampling algorithm, which reduces the gradient to the function value in the sense of expectation. Simply speaking, we consider the average of the function values in a small ball instead of the exact function value. The rigorous representation is to take the expectation of the function values along the following way as
\begin{equation}
\label{eqn: average-value}
\hat{J}(\pmb{u}_0) = \mathbb{E}_{\pmb{v}_0 \in \mathbb{B}^{d}} \left[ \nabla J(\pmb{u}_0 + \epsilon \pmb{v}_0) \right]
\end{equation}
where $\mathbb{B}^{d}$ is the unit ball in $\mathbb{R}^{d}$ and $\epsilon > 0$ is a small real number. According to Stokes' formula, we can derive the gradient of the expectation~\eqref{eqn: average-value} as
\begin{equation}
\label{eqn: approx-grad}
\nabla \hat{J}(\pmb{u}_0) = \mathbb{E}_{\pmb{v}_0 \in \mathbb{B}^d}\left[ \nabla J(\pmb{u}_0 + \epsilon \pmb{v}_0) \right] = \frac{d}{\epsilon} \cdot \mathbb{E}_{\pmb{v}_0 \in \mathbb{S}^{d-1}}\left[  J(\pmb{u}_0 + \epsilon \pmb{v}_0) \pmb{v}_0 \right].
\end{equation}
Following the expression~\eqref{eqn: approx-grad}, we can take the sample average to numerically approximate the gradient as 
\begin{equation}
\label{eqn: sample-grad}
\nabla \hat{J}(\pmb{u}_0) \approx \frac{d}{n\epsilon} \sum_{i=1}^{n}\left[  J(\pmb{u}_0 + \epsilon \pmb{v}_{0,i}) \pmb{v}_{0,i} \right],
\end{equation}
where $n$ is the number of samples and $\pmb{v}_{0,i},(i=1,\ldots,n)$ are the random variables identically sampled from the uniform distribution on the unit sphere $\mathbb{S}^{d-1}$. The rigorous Chernoff-type bound for the sample average with the exact gradient has been shown in~\citep[Section 3]{npg-30-263-2023}.

From the sample average of function values~\eqref{eqn: sample-grad}, we know that random variables $\pmb{v}_{0,i} ,(i=1,\ldots,n)$ are sampled independently from the (identical) uniform distribution on the unit sphere $\mathbb{S}^{d-1}$. In other words, for any two samples, $\pmb{v}_{0,i}$ and  $\pmb{v}_{0,j}$ with the indices satisfying $i \neq j$, there is no relationship between them, that is, every sample $\pmb{v}_{0,i} ,(i\in \{1,\ldots,n\})$ has no any influence with each other. Thus, when the modern parallel computation technique is considered, we can run the numerical model to obtain the values $ J(\pmb{u}_0 + \epsilon \pmb{v}_{0,i}) \pmb{v}_{0,i}$ for every $i \in \{1,\ldots,n\}$ simultaneously. Furthermore, if the resource of computation is unlimited, the time that we run $n$ times of numerical models can be reduced to that of running the model only once. Based on the current resource of computation, we realize the sampling algorithm to obtain the CNOPs, or the optimal precursors of the ZC model, the numerical model with dimension $O(10^{4}-10^{5})$, by use of the modern parallel computation technique, where the numerical performance is shown in~\Cref{sec: numerical-performance}.

\section{The Numerical Performance}
\label{sec: numerical-performance}

After the CNOP approach is imported to the ZC model~\citep{mu2007kind}, the adjoint method has always been the baseline algorithm in practice.  In this section, we show the numerical performance of the sampling algorithm by comparison with the adjoint method in the ZC model. The static spatial patterns of the optimal precursors with some measurement quantities and computation times are shown in~\Cref{subsec: cnops}, while the nonlinear time evolution behaviors of the optimal precursors and the corresponding Ni\~{n}o 3.4 SST anomaly index in~\Cref{subsec: cnops-evolution}

\subsection{The optimal precursors in the ZC model}
\label{subsec: cnops}
\begin{figure}[htb!]
\centering
\begin{subfigure}[t]{0.48\linewidth}
\centering
\includegraphics[scale=0.46]{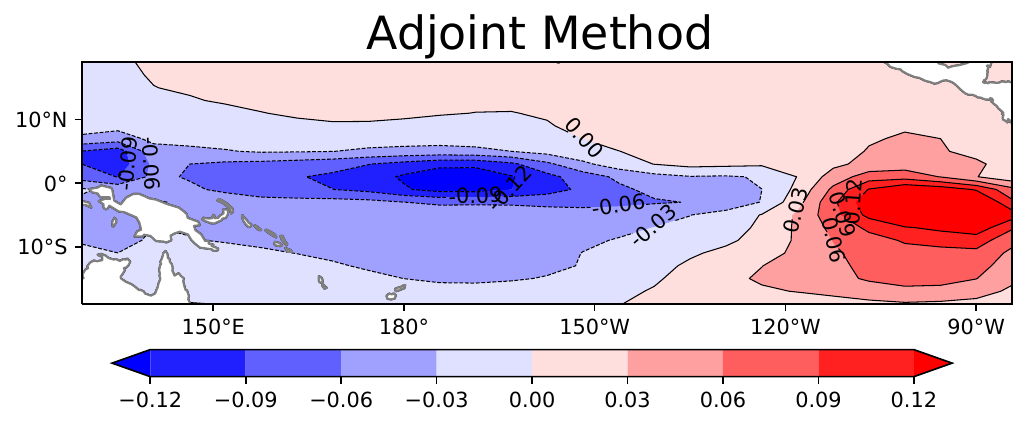}
\includegraphics[scale=0.46]{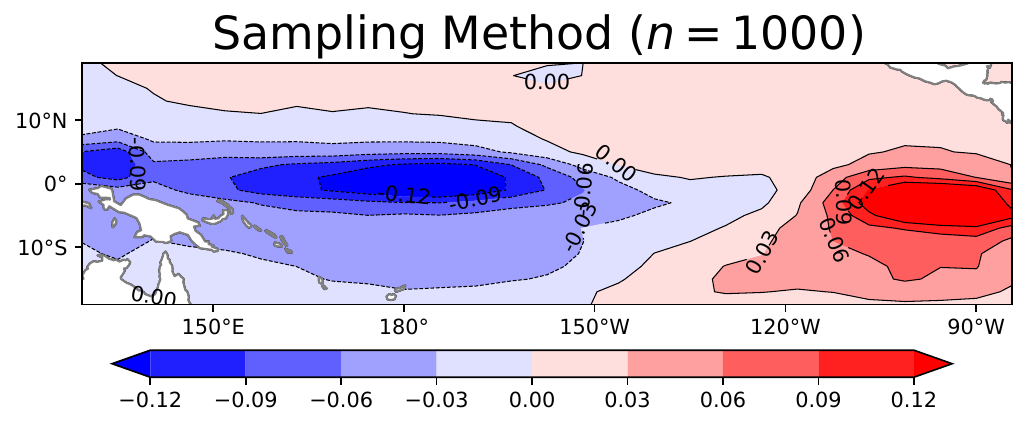}
\includegraphics[scale=0.46]{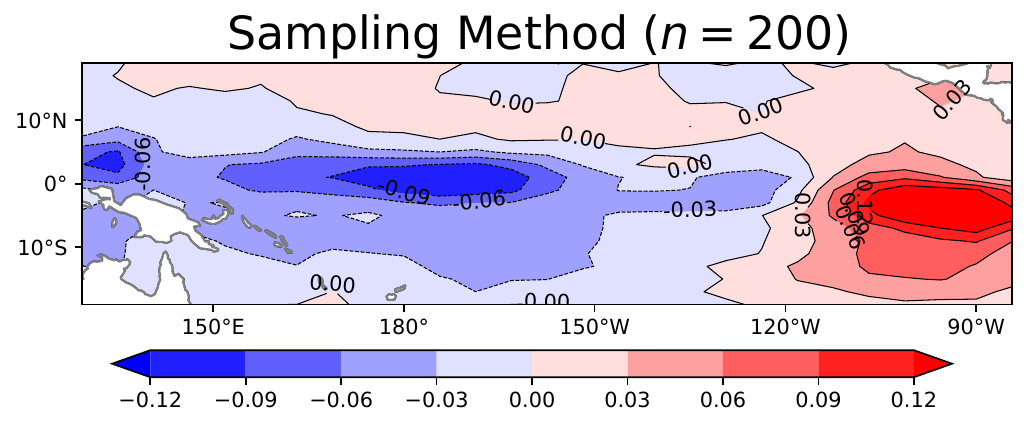}
\caption{SST Anomalies}
\end{subfigure}
\begin{subfigure}[t]{0.48\linewidth}
\centering
\includegraphics[scale=0.46]{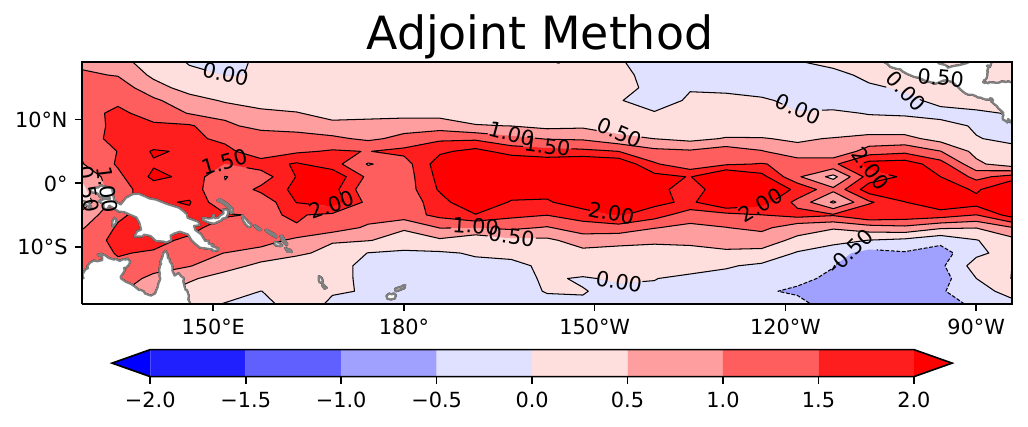}
\includegraphics[scale=0.46]{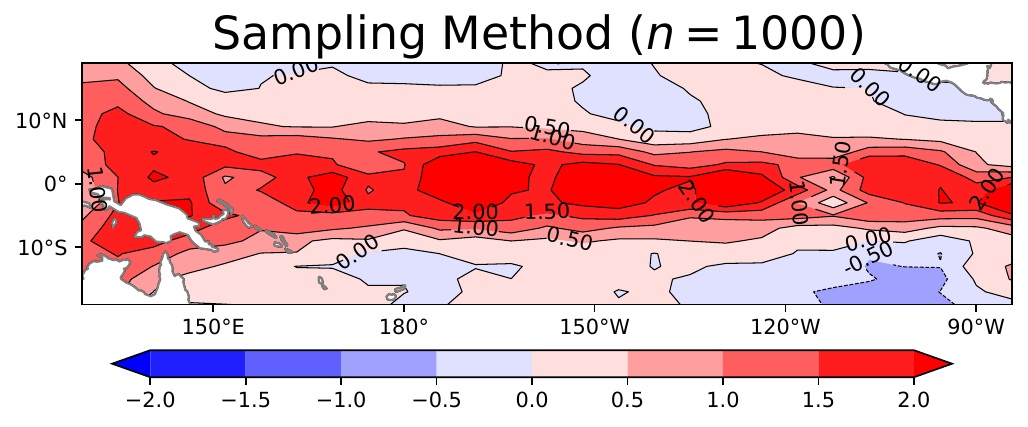}
\includegraphics[scale=0.46]{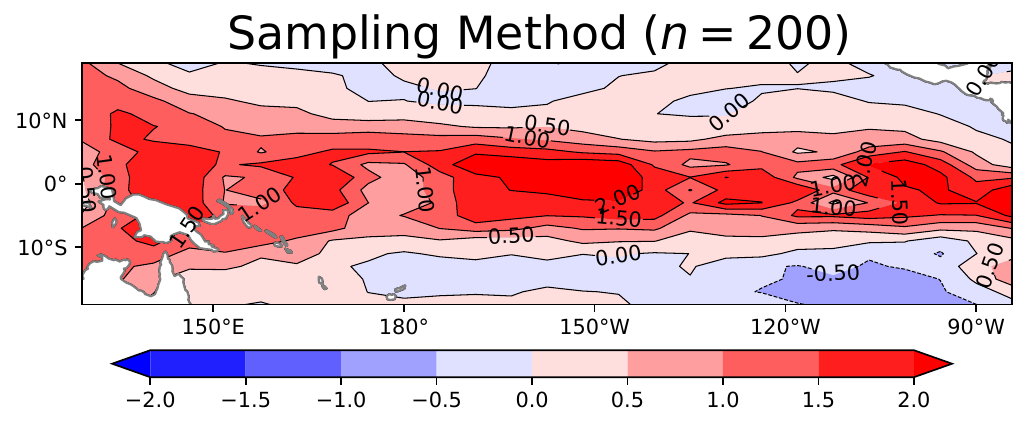}
\caption{Thermocline Depth Anomalies}
\end{subfigure}
\caption{The spatial patterns of the optimal precursors in terms of SST anomalies (Left Column) and thermocline depth anomalies (Right Column).  The prediction time is $9$ months. By the rows from top to bottom, the spatial patterns are obtained by the adjoint method, the sampling method with $n=100$ and that with $n=200$, respectively.} 
\label{fig: cnop-elnino}
\end{figure}

Recall the optimal precursors bringing about the El Ni\~{n}o event, which is obtained by the CNOP approach in~\citep{yu2009dynamics}. The spatial pattern in terms of SST anomalies is manifested as a large-scale zonal dipolar pattern, the warm pole of about $0.2^{\circ}\text{C}$ along the equator in the east Pacific and the cold one of about $0.2^{\circ}\text{C}$ in the central Pacific; while a basin-side deepening about $100 \text{m}$ along the equator is the character of that for thermocline depth anomalies. We reproduce the spatial patterns of the optimal precursors bringing about the El Ni\~{n}o event in~\Cref{fig: cnop-elnino}, the two pictures in the top row. When we take $1000$ samples to implement the sampling method, both the spatial patterns of the optimal precursors in terms of SST anomalies and thermocline depth anomalies are almost identical to that obtained by the adjoint method, which is shown in the middle row of~\Cref{fig: cnop-elnino}. Furthermore, when the number of samples is reduced from $1000$ to $200$, we can find from the two pictures in the bottom row of~\Cref{fig: cnop-elnino} that both the large-scale zonal dipolar pattern of SST anomalies and the basin-side deepening pattern of thermocline depth anomalies for the optimal precursors leading to the El Ni\~{n}o event can be still captured, even though there are some small deviations due to some noise. 


We have shown considerable similarities in the spatial patterns of the optimal precursors bringing about the El Ni\~{n}o event in~\Cref{fig: cnop-elnino}, which are obtained by the adjoint method, the sampling method with $n=1000$ and that with $n=200$. If these can be viewed as qualitative similarities,  we still need to verify the similarities of the optimal precursors from these numerical algorithms quantitatively.  The objective values of the optimal precursors are shown in~\Cref{tab:obj-vlaue}, where we can find that the one obtained by the adjoint method is $16.8441$ and that by the sampling method with $n=1000$ and $n=200$ are $16.6307$ and $15.4193$, respectively. The objective values obtained by the sampling algorithm look very close to that of the baseline adjoint method. Here, we can further show the similarities by taking the ratio between them.  If the objective value obtained by the adjoint method is taken as the numerator, we can find that the objective value obtained by the sampling method with $n=1000$ takes the percentage $98.73\%$, which manifests the objective values obtained by the two algorithms are almost identical. When the number of samples is reduced from $1000$ to $200$, the percentage that the objective value obtained by the sampling algorithm occupies decreases to $91.54\%$, which is still more than $90\%$ and shows quite high similarities.     

\begin{table}[htb!]
\centering
 \begin{tabular}{l|lll} 
   \hline
   \diagbox{Objective}{Methods}           & Adjoint           &Sampling ($n=1000$) & Sampling ($n=200$)                       \\ 
   \hline
   Values ($\|\pmb{T}(\tau)\|$)         & $16.8441$     & $16.6307$                  & $\mathbf{15.4193}$                     \\
   Percentage                              & $100\%$        & $98.73\%$                  & $\mathbf{91.54\%}$                  \\
    \hline
 \end{tabular}
 \caption{The objective values of the optimal precursors and the percentages over that computed by the adjoint method. Bold emphasizes the high efficiency of the sampling method with $n=200$.} 
 \label{tab:obj-vlaue}
\end{table}

Both the spatial patterns and the objective values show that the optimal precursor that we implement the sampling algorithm by taking only $200$ samples to obtain is very similar to that obtained by the baseline adjoint algorithm. To show the efficiency of the sampling method, we still need to compare the computation time. Based on the description that the sampling method can be implemented with parallel computation in~\Cref{subsec: parallel-computation}, we realize them and show the computation times in~\Cref{tab:comp-time}. It needs to take about $50$ iterations by implementing both the adjoint method and the sampling method to get the optimal precursors. On the supercomputer, the adjoint method takes about $15s$, that is, about $0.3s$ per iteration; while the sampling method takes about $3s$ when the implementation is under the parallel computation, that is, about $0.06s$ per iteration. Furthermore, when the sampling method is implemented, we avoid running the numerical model reversely such that the computation time is shortened to $1/5$. Without any doubt, the computation that is reduced must be implemented by parallel computation. However, based on the current resource of computation, it is available for us to implement the sampling method under the parallel computation to obtain the optimal precursors by the use of the CNOP approach in the ZC model, more generally, the numerical model with dimension $O(10^{4}-10^{5})$.
\begin{table}[htb!]
\centering
 \begin{tabular}{lll} 
   \hline
   Methods                                                                & Adjoint                           &Sampling (Parallel Computation)  \\ 
   \hline
   Computation Time ($50$ Iterations)                    & $\approx 15s$               & $\approx 3s$                                \\
   Computation Time per Iteration                          & $\approx 0.3s$             & $\approx 0.06s$                              \\
    \hline
 \end{tabular}
 \caption{The comparison of computation times between the adjoint method and the sampling method under the parallel computation. Run the Fortrun code on the CPU: Intel\textsuperscript{\tiny\textregistered} Xeon\textsuperscript{\tiny\textregistered} Gold 6132 Processor, 19.25M Cache, 2.60 GHz.} 
 \label{tab:comp-time}
\end{table}

\subsection{The nonlinear time evolution behavior of the optimal precursors}
\label{subsec: cnops-evolution}

Based on the CNOP approach,  the statically spatial patterns of the optimal precursors of ENSO events are in terms of both SST anomalies and thermocline depth anomalies. In~\Cref{subsec: cnops}, we have shown the high efficiency of the sampling algorithm by the comparison with that obtained by the baseline adjoint method as well as the computation times. However, we still need to study the dynamic behaviors of the ENSO events to predict the potential impacts, where a great way is to only monitor the nonlinear evolution of SST anomalies.   

\begin{figure}[htbp!]
\centering
\begin{subfigure}[t]{0.325\linewidth}
\centering
\includegraphics[scale=0.31]{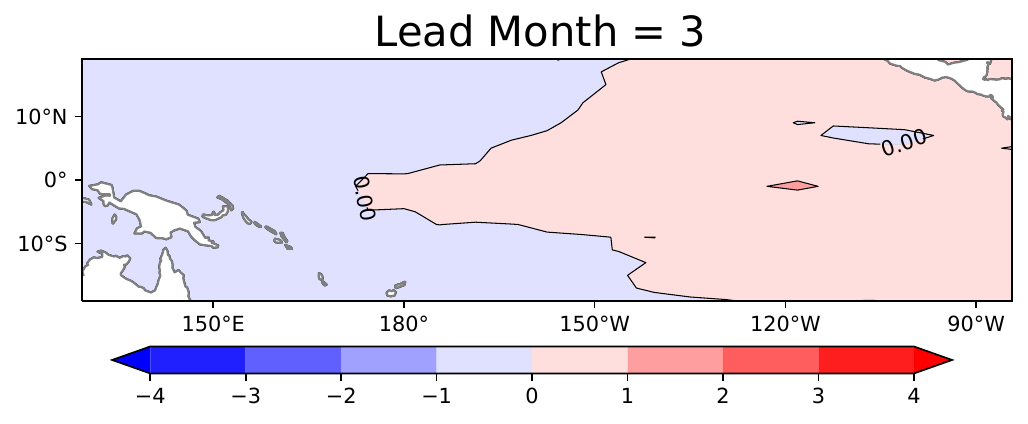}
\includegraphics[scale=0.31]{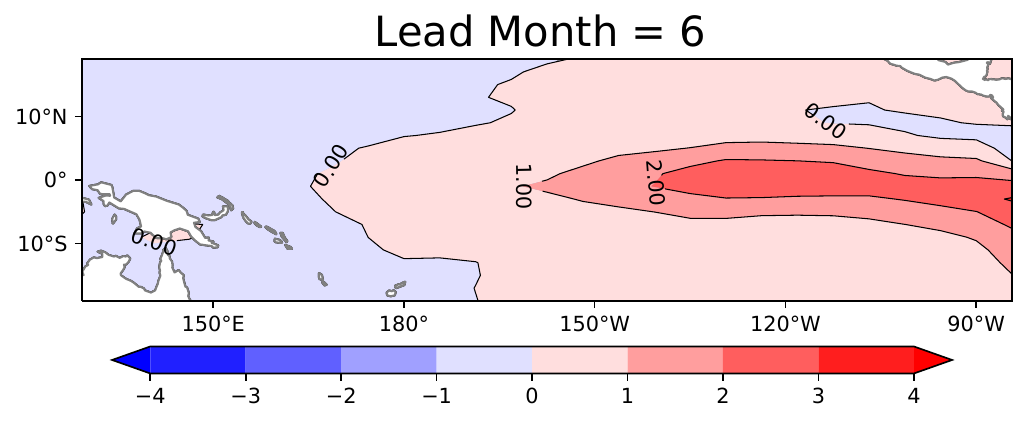}
\includegraphics[scale=0.31]{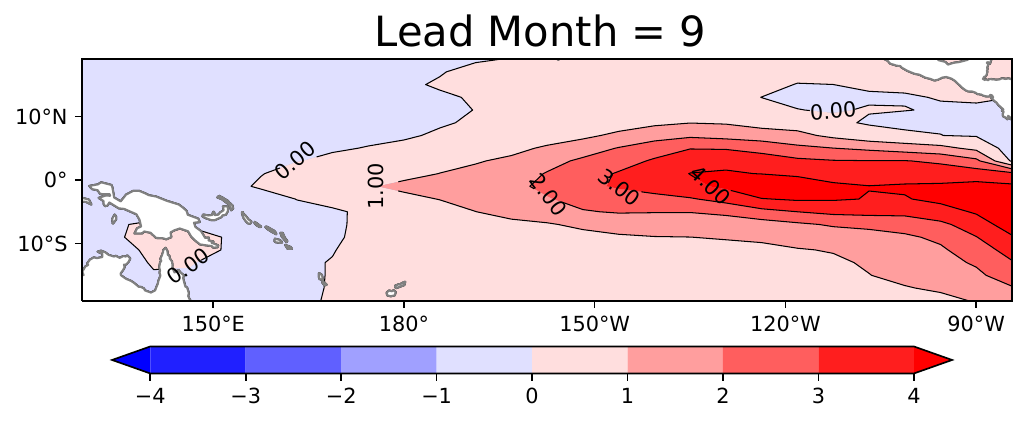}
\includegraphics[scale=0.31]{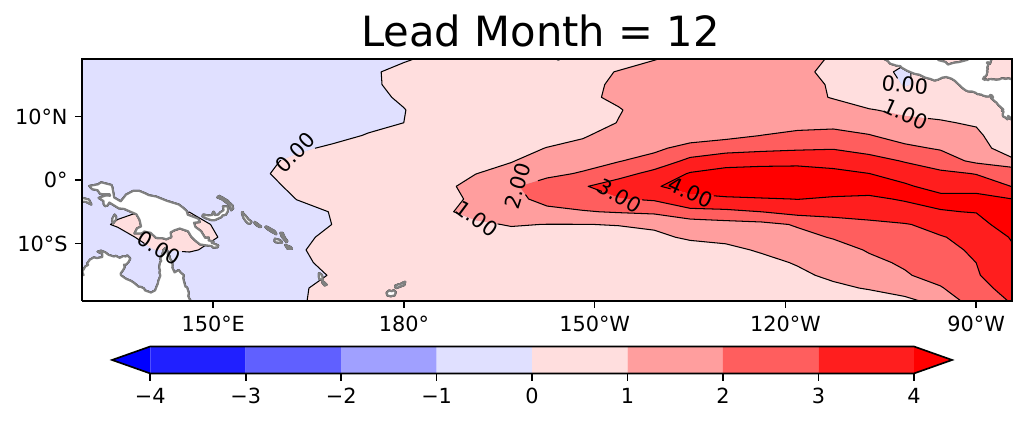}
\caption{Adjoint Method}
\end{subfigure}
\begin{subfigure}[t]{0.325\linewidth}
\centering
\includegraphics[scale=0.31]{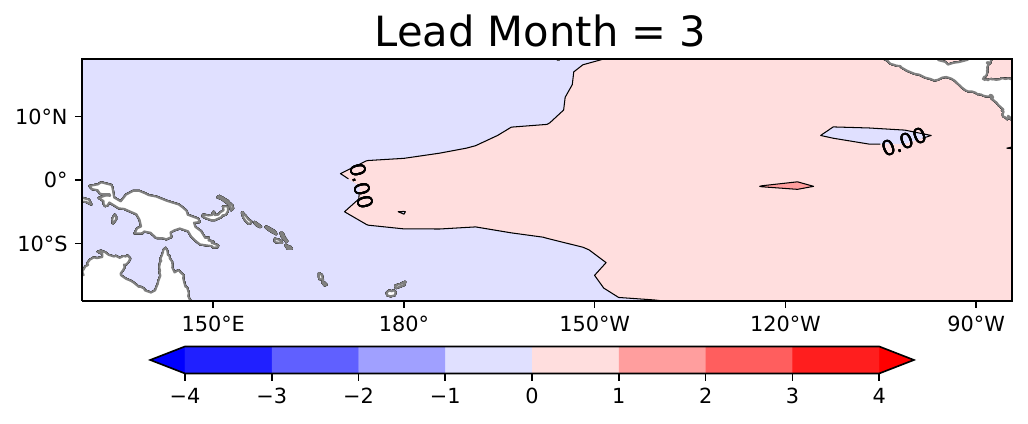}
\includegraphics[scale=0.31]{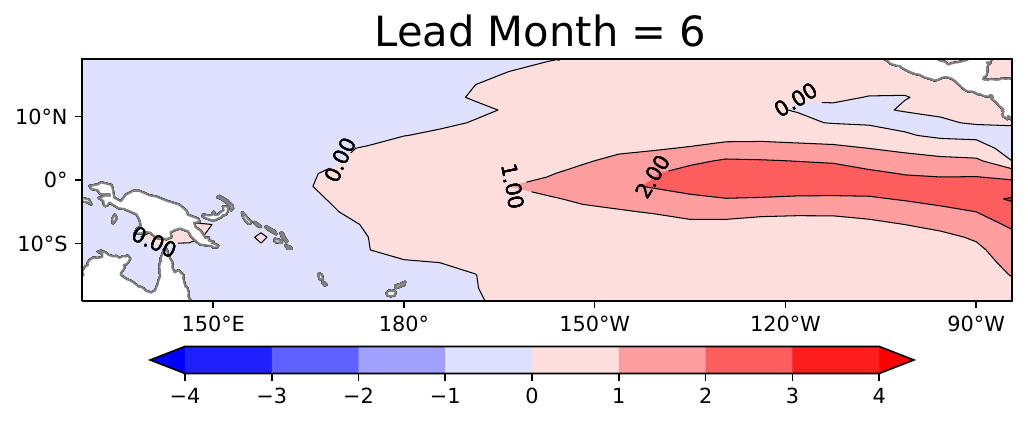}
\includegraphics[scale=0.31]{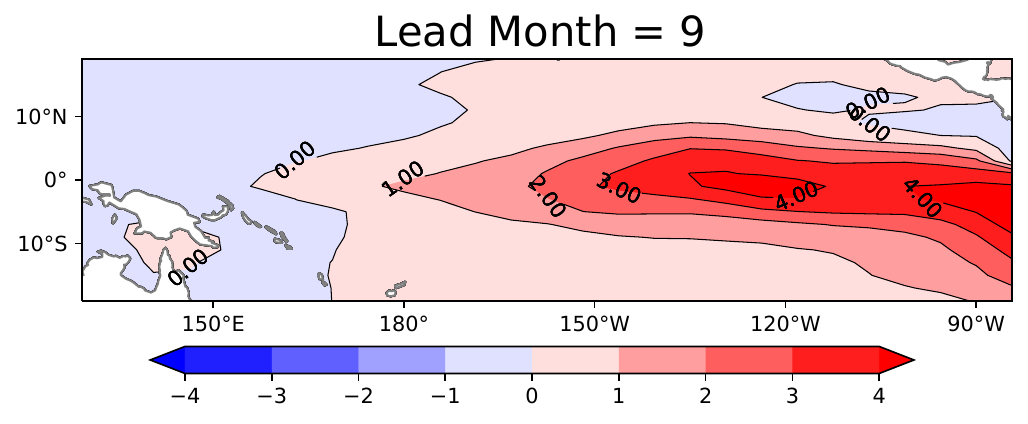}
\includegraphics[scale=0.31]{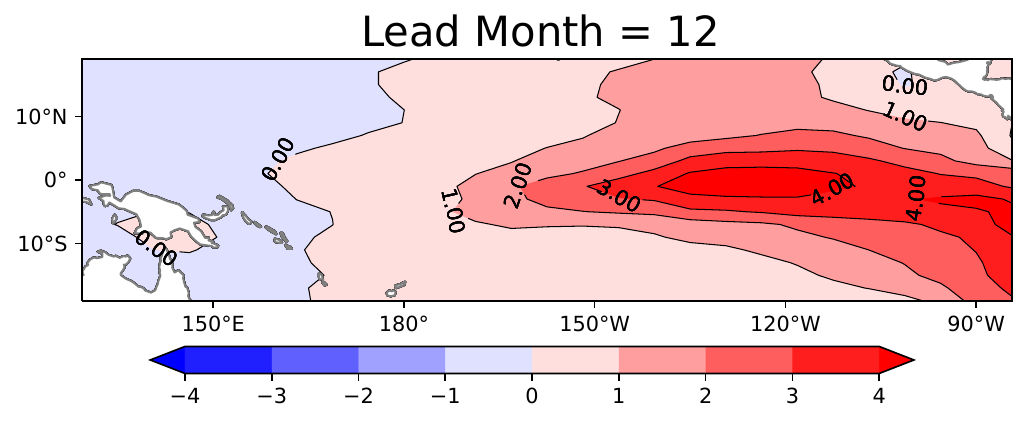}
\caption{Sampling Method ($n=1000$)}
\end{subfigure}
\begin{subfigure}[t]{0.325\linewidth}
\centering
\includegraphics[scale=0.31]{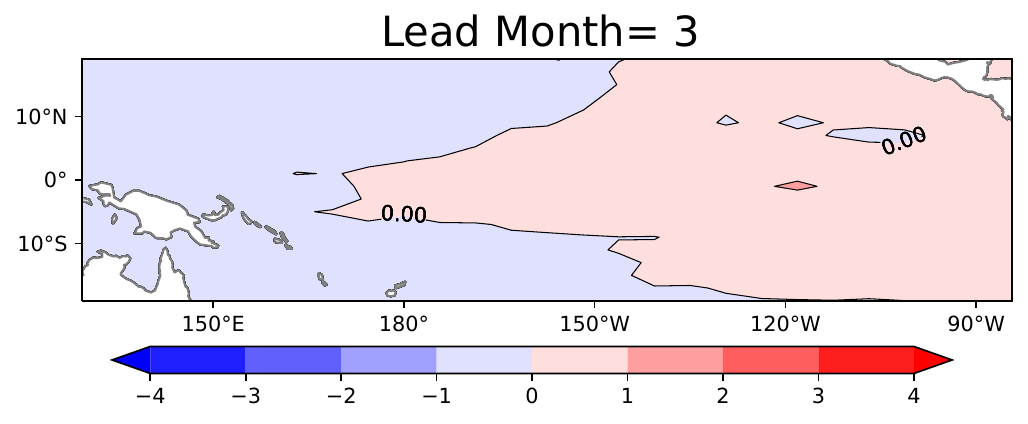}
\includegraphics[scale=0.31]{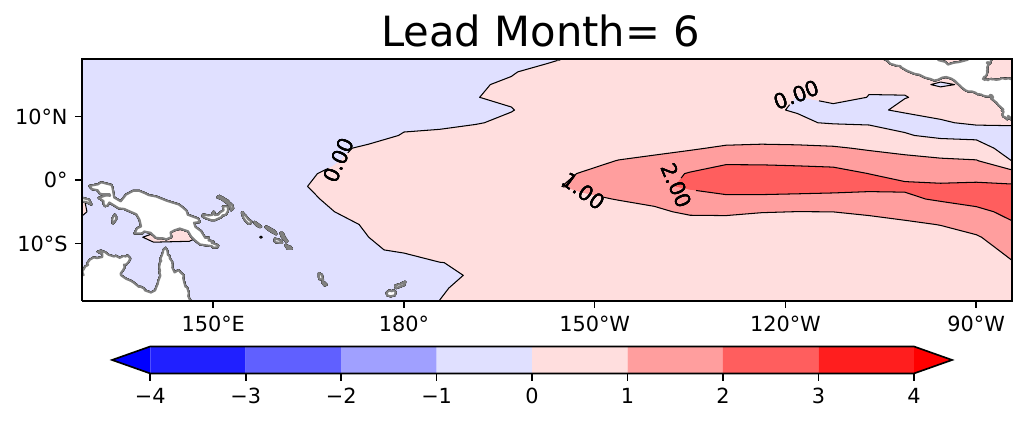}
\includegraphics[scale=0.31]{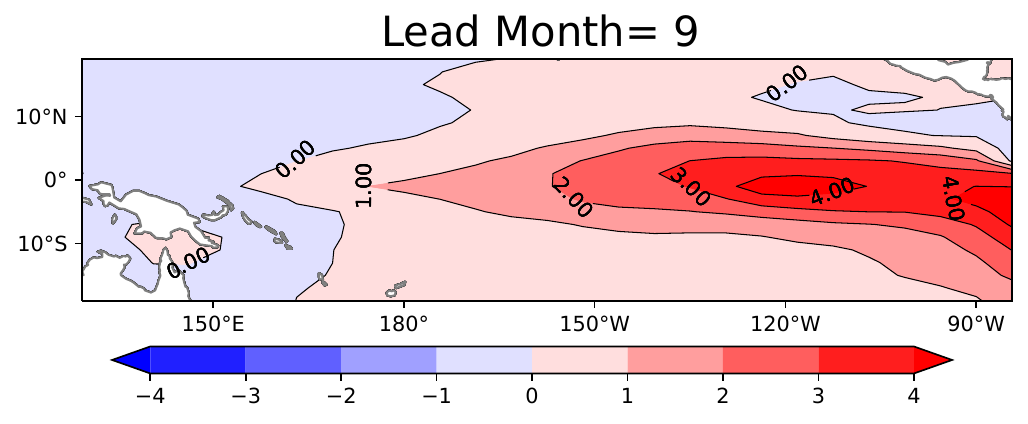}
\includegraphics[scale=0.31]{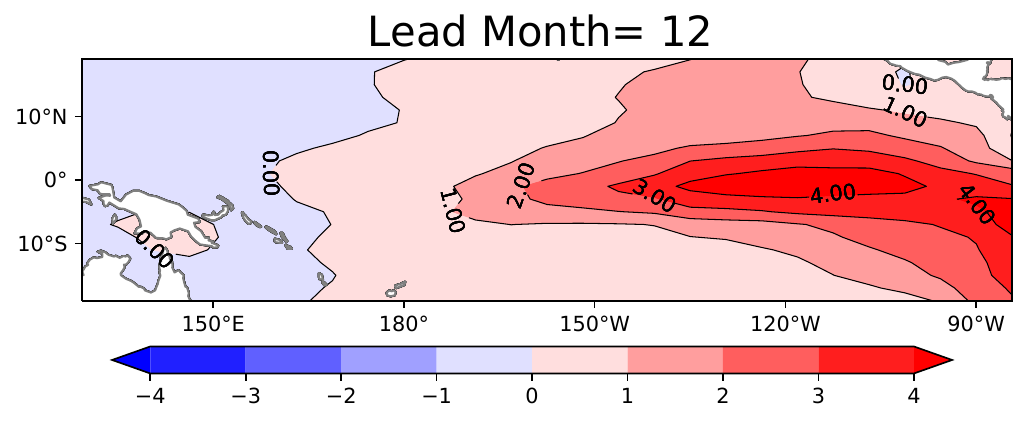}
\caption{Sampling Method ($n=200$)}
\end{subfigure}
\caption{The spatial patterns of the nonlinear time evolution of the optimal precursors in terms of SST anomalies.} 
\label{fig: evolution-elnino}
\end{figure}

Recall the nonlinear time evolution of SST anomalies simulated by the coupled ocean-atmosphere ZC model shown in~\citep{yu2009dynamics}, where the optimal precursor, or the CNOP, is obtained by the adjoint method. By adding the initial optimal precursors to the climatological mean equilibrium state, we run the ZC model to reproduce the EP-type El Ni\~{n}o phenomenon in the left column of~\Cref{fig: evolution-elnino}, where we can observe that the warm phase in the east Pacific along the equator is intensified gradually with the season evolution in one year, that is, the lead time is set as $3$, $6$, $9$, and $12$ months, respectively. More concretely, in the east Pacific along the equator, the region of the warm phase is gradually enlarged and the SST anomalies are raised up sharply from about $0^{\circ}\text{C}$ to $8^{\circ}\text{C}$.  In the right two columns of~\Cref{fig: evolution-elnino}, we show the spatial patterns in terms of the nonlinear time evolution of SST anomalies, where the initial condition starts from the climatological mean equilibrium state added by the initial optimal precursors obtained by the sampling methods with $n=1000$ and $n=200$. By taking a comparison between the spatial patterns shown from the left to the right in~\Cref{fig: evolution-elnino}, the nonlinear time evolution behaviors of the initial optimal precursors are also remarkably similar to each other with the change of seasons. Even though the number of samples is reduced to $200$, we can still find that the spatial patterns in terms of the seasonal evolution of SST anomalies are almost consistent with the baseline one starting from initial optimal precursors obtained by the adjoint method.

\begin{figure}[htbp!]
\centering
\includegraphics[scale=1.30]{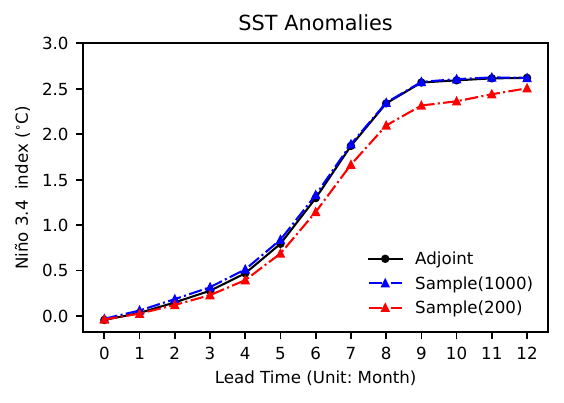}
\caption{The nonlinear time evolution of Ni\~{n}o 3.4 SST anomaly index within a model year. } 
\label{fig: nino-index}
\end{figure}

Based on the nonlinear time evolution of SST anomalies simulated in~\Cref{fig: evolution-elnino}, we have shown qualitatively the similarities of the dynamical behaviors of the initial optimal precursors obtained by both the baseline adjoint method and the sampling method. Nevertheless, we still need to show the similarities quantitatively for the dynamical evolution of SST anomalies from the three kinds of initial optimal precursors. Currently, the main variable that is considered from the ENSO forecasts of the coupled climate models is the Ni\~{n}o 3.4 SST anomaly index, which is used by the National Climate Centre (NCC) in Australia to classify ENSO conditions. Here, we show that the Ni\~{n}o 3.4 SST anomaly indices change nonlinearly along the time evolution line within a model year in~\Cref{fig: nino-index},
where the three dynamical curves generated by these proposed algorithms are quite close to each other. Furthermore, we can observe in~\Cref{fig: nino-index} that the dynamical curve of the Ni\~{n}o 3.4 SST anomaly index starting from the initial optimal precursor obtained by the sampling method with $n=1000$ almost coincides with the baseline one from the initial optimal precursor obtained by the adjoint method. When the number of samples is reduced from $1000$ to $200$, some small deviations appear in the dynamical curve of the Ni\~{n}o 3.4 SST anomaly index. 
\begin{figure}[htbp!]
\centering
\includegraphics[scale=1.30]{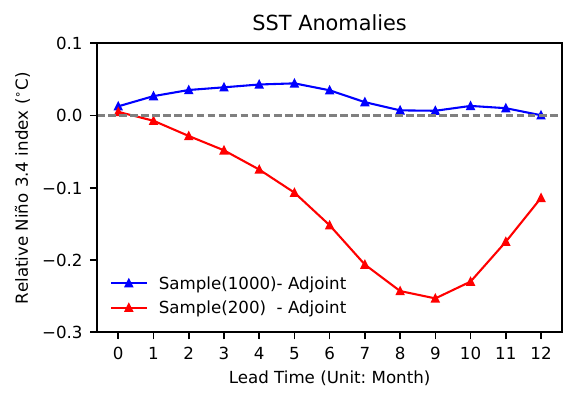}
\caption{The nonlinear time evolution of relative Ni\~{n}o 3.4 SST anomaly index within a model year. (the Ni\~{n}o 3.4 SST anomaly index obtained by the sampling method minus that by the baseline adjoint method)} 
\label{fig: nino-index-difference}
\end{figure}
Thus, it is necessary for us to quantify these derivations such that we can study the accuracy of the Ni\~{n}o 3.4 SST anomaly index by implementing the sampling algorithm to approximate that generated by the adjoint method when the number of samples is reduced from $1000$ to $200$. Taking the Ni\~{n}o 3.4 SST anomaly index generated by the adjoint method as a basis, we compute the relative Ni\~{n}o 3.4 SST anomaly index, that is, the difference of the Ni\~{n}o 3.4 SST anomaly indices from between the baseline adjoint method and the sampling method in~\Cref{fig: nino-index-difference}. Here, we can find that the relative  Ni\~{n}o 3.4 SST anomaly index from the sampling method with $n=1000$ takes the characteristic scale with $\mathrm{O}(10^{-2})$, while that from the sampling method with $n=200$ is $\mathrm{O}(10^{-1})$. In other words, when we implement the sampling method by reducing the number of samples from $n=1000$ to $n=200$, the relative Ni\~{n}o 3.4 SST anomaly index is degraded from $\mathrm{O}(10^{-2})$ to $\mathrm{O}(10^{-1})$, which quantitatively manifests that the accuracy of the Ni\~{n}o 3.4 SST anomaly index is loosened up to an order of magnitude. However, if we take the comparison with the Ni\~{n}o 3.4 SST anomaly index, whose characteristic scale is $\mathrm{O}(1)$, the numerical errors by reducing the number of samples from $1000$ to $200$ are still too small to influence the nonlinear time evolution of SST anomalies.

\section{Summary and discussion}
\label{sec: summary-discussion}

Based on the state-of-the-art statistical machine learning techniques,  the sampling method to compute the CNOPs is proposed in~\citep{npg-30-263-2023}. In this paper, we successfully implement the sampling method to obtain the initial optimal precursors in the realistic and predictive ZC model, more generally, the numerical model with dimension $\mathrm{O}(10^{4}-10^{5})$. The sampling method with fewer samples can achieve consistent performance as the adjoint method in the numerical experiments, regardless of the statically spatial patterns and the dynamical nonlinear time evolution behaviors as well as the corresponding quantities and indices. Based on the key characteristic that the samples are independently and identically distributed, we can implement the sampling method by the modern parallel computation technique to eliminate running the numerical model reversely such that the computation time is shortened extremely, about $1/5$. Indeed, we take $60$ samples to implement the sampling algorithm which can reproduce the numerical performance of the initial optimal precursors obtained by the baseline adjoint method. However, the numerical results are unstable. Specifically, one in four times that we run the coupled ocean-atmosphere ZC model will lead to a correct numerical performance. Besides, by the use of the CNOP approach in the coupled ocean-atmosphere ZC model, we can obtain the other kind of optimal precursors,  which leads to the La Ni\~{n}a event. However, due to the deficiency of the original ZC model in~\citep{zebiak1987model}, a warm tendency of the Ni\~{n}o 3.4 SST anomaly index will appear after it decreases to the coldest point for the La Ni\~{n}a event, which is shown in~\citep{duan2008decisive}. In our numerical experiments, the numerical performance based on the optimal precursors leading to the La Ni\~{n}a event can be also obtained. Thus, these numerical experiments are not representative such that we neglect to show their numerical performance in the paper.

For a realistic global climate system model (GSCM) or atmosphere-ocean general circulation model (AOGCM), it is often impractical to develop the adjoint model, so the sampling method provides a probable way of computing the CNOPs to investigate its predictability. An interesting direction for further research is to investigate the CNOPs computed by the sampling method in the numerical models that are used in realistic prediction and forecast, such as the Weather Research and Forecasting (WRF) Model, a state-of-the-art mesoscale numerical weather prediction system for operational forecasting applications. In addition, the traditional data assimilation is based on the development of the adjoint model~\citep{kalnay2003atmospheric}. In this paper, our numerical experiments make the four-dimensional variational (4D-Var) data assimilation  technique become available possibly on the coupled climate system models as well as the parameterization models. Therefore, it is valuable and thrilling to implement the sampling method to process 4D-Var data assimilation in realistic systems, such as the Flexible Global Ocean-Atmosphere-Land System (FGOALS)-s2~\citep{wu2018enoi} for decadal climate prediction.

\section*{Acknowledgments}
{\small This work was supported by Grant No.12241105 of NSFC and Grant No.YSBR-034 of CAS.}

{\small
\bibliographystyle{abbrvnat}
\bibliography{sigproc}}


\end{document}